\documentclass[prd,eqsecnum,preprintnumbers,showpacs]{revtex4}
\usepackage{amsfonts}
\usepackage{amsmath}
\usepackage{amssymb}
\usepackage{natbib}
\usepackage{graphicx}
\usepackage{subfigure}
\usepackage{psfrag}

\DeclareMathOperator{\sinc}{sinc}

\begin{document}
\title[Response of interferometric detectors]{The response of interferometric gravitational wave detectors}

\author{Lee Samuel Finn}
\altaffiliation{Center for Gravitational Wave Physics, Department of Physics, Department of Astronomy and Astrophysics}
\affiliation{The Pennsylvania State University, University Park, Pennsylvania 16802, USA}

\date{\today}

\begin{abstract}
The derivation of the response function of an interferometric gravitational wave detector is a paradigmatic calculation in the field of gravitational wave detection. Surprisingly, the standard derivation of the response wave detectors makes several unjustifiable assumptions, both conceptual and quantitative, regarding the coordinate trajectory and coordinate velocity of the null geodesic the light travels along. These errors, which appear to have remained unrecognized for at least 35 years, render the ``standard'' derivation inadequate and misleading as an archetype calculation. Here we identify the flaws in the existing derivation and provide, in full detail, a correct derivation of the response of a single-bounce Michelson interferometer to gravitational waves, following a procedure that will always yield correct results; compare it to the ``standard'', but incorrect, derivation; show where the earlier mistakes were made; and identify the general conditions under which the ``standard'' derivation will yield correct results. By a fortuitous set of circumstances, not generally so, the final result is the same in the case of Minkowski background spacetime, synchronous coordinates, transverse-traceless gauge metric perturbations, and arm mirrors at coordinate rest. 
\end{abstract}


\maketitle

\section{Introduction}\label{sec:intro}
The derivation of the response function of an interferometric gravitational wave detector is a paradigmatic calculation in the field of gravitational wave detection. The same general derivation appears, almost without variation, in reference books \cite{saulson:1994:foi}, pedagogical and review articles \cite{hellings:1983:efg,saulson:1997:ilw,schutz:2001:gws}, general relativity texts \cite{schutz:1990:fci}, proposals for experimental facilities and technical reports \cite{weiss:1972:ecb,linsay:1983:sol}, and other highly-cited articles from the primary literature (e.g., \cite{schutz:1987:apo,cornish:2001:smt,cornish:2003:lrf,baskaran:2004:cog}). 
Derivations of the response come in two flavors. By far the most popular is  based on the vanishing proper length along a null geodesic separating the ends of each interferometer arm. The second flavor of derivation, whose validity is restricted to wavelengths much larger than the size of the detector, uses the equation of geodesic deviation to calculate the changing separation between the mirrors that bound the detectors arms. 

Surprisingly, both these derivations of the detector response function are conceptually and methodologically incorrect. Their flaws appear to have gone undetected for at least 35 years \footnote{At a late stage in this work I became aware of a related investigation by Rakhmanov (in preparation) and suspicions regarding the correctness of these expressions by Whelan in a footnote to \cite{whelan:2007:hct}.}, going back to the original proposals for constructing gravitational wave detectors of this type \cite{weiss:1972:ecb,forward:1978:wlg}. Perhaps equally surprising, a correct derivation of the response, which exposes the several errors made in the classic derivations, also finds that the combination of errors cancel in the special case of gravitational wave perturbations about flat spacetime in synchronous coordinates and transverse-traceless (TT) gauge. In more general circumstances, however, the errors are significant and using the standard derivations as a template for similar calculations in these more general circumstances will lead to order unity fractional errors. 

In Sec.~\ref{sec:classic} we recap the standard derivations of the response of an interferometric gravitational wave detector and highlight the conceptual errors that invalidate them. In Sec.~\ref{sec:correct} we provide, in full detail, a correct derivation of the response of a single-bounce Michelson interferometer to gravitational waves. We take special care to justify every step in the derivation and highlight where the intermediate results in the standard calculation are quantitatively incorrect. Finally, in Sec.~\ref{sec:discuss} we discuss the special circumstances that lead the intermediate errors to cancel when the entire calculation is assembled and applied to find the detector response. 

\section{``Classic'' Derivations of the Detector Response}\label{sec:classic}

\subsection{Physical observable}\label{sec:observable}
\begin{figure}
\begin{center}
\includegraphics[width=4.0in]{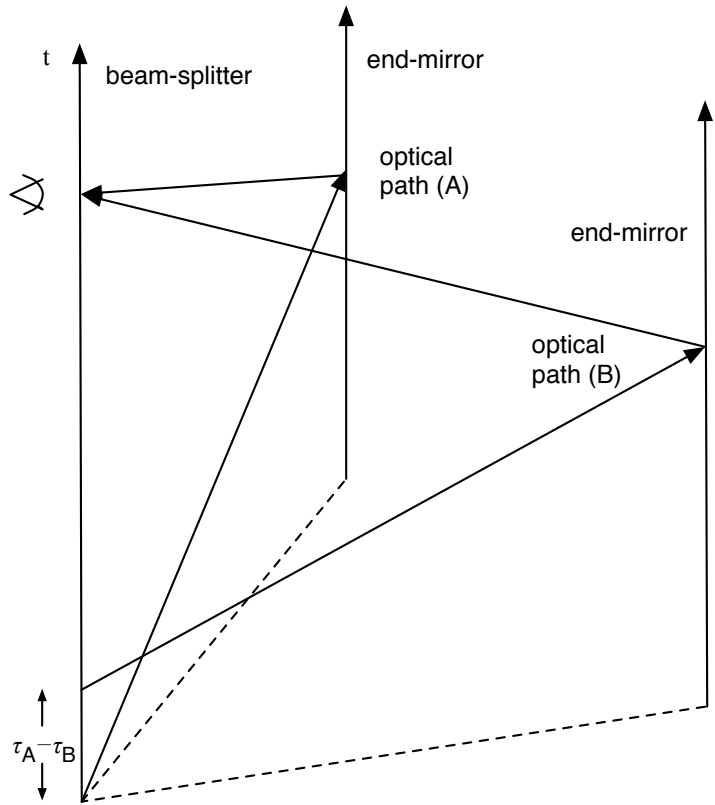}
\caption{Spacetime diagram of the optical paths in a simple Michelson interferometer. The dashed lines lie in a constant time hypersurface. The observer measures the difference in the phase of light that has traveled along the two different optical paths $(\mathsf{A})$ and $(\mathsf{B})$. In the case of a common, monochromatic light source at the beam splitter this difference is the difference in phase of the light source at times $t-\tau_{\mathsf{A}}$ and $t-\tau_{\mathsf{B}}$.}\label{fig:ifo}
\end{center}
\end{figure}

Consider a simple Michelson laser interferometer, shown schematically in Fig.\ \ref{fig:ifo}. By various devices (cf.~\cite{saulson:1994:foi}) one can measure the quantity $(\Delta\phi(t))\mod(2\pi)$, where $\Delta\phi$ is the phase difference $\phi_{(\mathsf{A})}(t)-\phi_{(\mathsf{B})}(t)$ in the light arriving at the interferometer output port along optical paths $(\mathsf{A})$ and $(\mathsf{B})$. The light arriving and recombined at the beam splitter along these paths originated at the beam splitter at earlier times $t-\tau_{\mathsf{A}}$ and $t-\tau_{\mathsf{B}}$, so that 
\begin{equation}
\Delta\phi(t) = \phi_{(\mathsf{A})}(t)-\phi_{(\mathsf{B})}(t) = \phi_0(t-\tau_{\mathsf{A}}) - \phi_0(t-\tau_{\mathsf{B}})
\end{equation}
where $\phi_0$ is the phase of the light leaving the beam splitter as a function of time. If the light source is, e.g., a laser with frequency $\nu$ and we choose coordinates such that $t$ is proper time along the beam splitter's world line then $\Delta\phi(t)=2\pi\nu(\tau_{\mathsf{B}}-\tau_{\mathsf{A}})$. In this way a laser interferometer is sensitive to gravitational wave perturbations in the difference in time, measured at the beam splitter, required for light to propagate along the optical paths $(\mathsf{B})$ and $(\mathsf{A})$. To determine the interferometer response, then, is to determine the gravitational wave perturbation to the difference in these light travel times. 

\subsection{Derivation from vanishing proper length of null paths}

The most popular derivation of the interferometric gravitational wave detector response is based on the vanishing proper length along a null path separating the two ends of each interferometer arm. Examples of this derivation are found in reference books 
\cite{saulson:1994:foi}, 
%
% textbooks \cite{schutz:1990:fci}, 
%
review articles \cite{hellings:1983:efg},
pedagogical articles and summer school proceedings \cite{saulson:1997:ilw,schutz:2001:gws}, 
proposals and technical reports \cite{weiss:1972:ecb,linsay:1983:sol}, 
%
% detector descriptions \cite{forward:1978:wlg}, and
%
and other highly cited articles from the primary literature \cite{forward:1978:wlg,schutz:1987:apo,cornish:2001:smt,cornish:2003:lrf,baskaran:2004:cog}.
The starting point for this derivation is the invariant interval in the perturbed spacetime evaluated in transverse-traceless (TT) gauge \cite{misner:1973:g} for a null path connecting the beam splitter and an end mirror:
\begin{equation}\label{eq:beginClassic}
0 = g_{\mu\nu}{dx^\mu}{dx^\nu} = -dt^2 + \delta_{ij}dx^idx^j + h_{ij}dx^idx^j.
\end{equation}
This expression is rewritten as 
\begin{align}
\int dt &= \int d\lambda\,\sqrt{\frac{dx^i}{d\lambda}\frac{dx^j}{d\lambda}\left(\delta_{ij}+h_{ij}\right)}\nonumber\\
&= \int d\lambda\,
\sqrt{\delta_{ij}\frac{dx^i}{d\lambda}\frac{dx^j}{d\lambda}}
\left(
1+\frac{1}{2}\frac{h_{kl}\frac{dx^k}{d\lambda}\frac{dx^l}{d\lambda}}{\delta_{pq}\frac{dx^p}{d\lambda}\frac{dx^q}{d\lambda}}
\right)\label{eq:rhs}
\end{align}
where the perturbation $h_{ij}$ is taken to be small and the integration path is along the null geodesic. The integral on the left is $\Delta t$, the light travel time along the null geodesic path between the beam splitter and an end mirror. To evaluate the right-hand side integral one introduces the \emph{unperturbed} spatial trajectory of an electromagnetic phase front as it propagates along a null-ray from the beam splitter to the end mirror at $z^j$ (cf.~e.g., 
\cite[Eq.~2.2.15]{saulson:1994:foi}, 
\cite[Eq.~1--3]{hellings:1983:efg},
\cite[Eqs.~2.3--2.4]{saulson:1997:ilw},
\cite[Eqs.~2.16--2.18]{schutz:2001:gws},
\cite[pg.~2]{weiss:1972:ecb},
\cite[pgs.~III-1--III-3]{linsay:1983:sol}
\cite[Eqs.~35--39]{baskaran:2004:cog}, 
\cite[Eqs.~A2--A4]{cornish:2001:smt},
\cite[Eqs.~4--5]{cornish:2003:lrf}):
\begin{subequations}\label{eq:unperturb}
\begin{align}\label{eq:trajParm}
t &= \lambda,\qquad{} x^j = \lambda \zeta^j \\
1 &= \zeta^j \zeta^k \delta_{jk} .
\end{align}
\end{subequations}
With this substitution the integral on the right becomes
\begin{equation}\label{eq:L}
\int_0^L d\lambda\,\left(1+\frac{1}{2}h_{kl}(t,x^j)\zeta^k\zeta^l\right)  = L + \frac{1}{2}\int_0^L d\lambda\,h_{kl}(t,x^j)\zeta^k\zeta^l.  
\end{equation}
When the gravitational wave period $f^{-1}$ is much less than $L$ (i.e., $fL\ll1$) the perturbation $h_{ij}$ is constant along the path and the second integral simplifies to $Lh_{ij}\zeta^i\zeta^j/2$, where $\zeta^i$ is the unit vector coordinate direction from the beam splitter to the end mirror. The result is 
\begin{align}
\Delta t &= L + \frac{L}{2} h_{ij}\zeta^i\zeta^j. 
\end{align}
What is observed at the beam splitter is proportional to the difference in the light travel time, from the beam splitter to an end mirror and back, along each arm, or
\begin{equation}\label{eq:endClassic}
\delta(\Delta t) = L h_{ij}\left(\zeta^i\zeta^j - \xi^i\xi^j\right),
\end{equation}
where $\xi^i$ is the unit vector along the second interferometer arm, also of length $L$, and one assumes a single round-trip of the light along each arm. The generalization of this result to a multibounce delay line, Fabrey-Perot cavity interferometer, and/or a light storage time greater than the wave period is straightforward. 

A careful reader may question the evaluation of the right-hand side integral in Eq.\ (\ref{eq:rhs}) along the unperturbed null-geodesic trajectory described in Eq.\ (\ref{eq:unperturb}) leading to the right-hand side of Eq.\ (\ref{eq:L}). In particular, use of the unperturbed spatial trajectory of the null-ray ignores $\mathcal{O}(h)$ corrections to the coordinate trajectory $dx^j/dt$ and the coordinate $x^j$. Why these  two $\mathcal{O}(h)$ corrections should be ignored in the evaluation of the $\mathcal{O}(1)$ contribution to the integrand in Eq.\ (\ref{eq:rhs}) is not apparent and never justified. That the spacetime trajectory is geodesic is not relevant here: the integral in question is not over the spacetime path and, in any event, the coordinate trajectories of the null geodesic connecting the beam splitter and end mirror world lines in the perturbed and unperturbed spacetimes have different end-points ---  indeed, it is exactly the difference $L-\tau$ that we desire to calculate.

In fact, the standard derivation is flawed on exactly this point. Naively following this ``classic'' derivation in circumstances where it may, on the basis of all standard presentations that have been examined, be thought to apply (e.g., availability of TT gauge and coordinate stationary mirrors, which are the usually remarked-upon circumstances) will, in general, yield incorrect results.  As discussed in Sec.\ \ref{sec:discuss}, for perturbations of a restricted set of spacetimes in special coordinates the errors resulting from the use of the unperturbed trajectory cancel; however, the reasons for this cancellation are nontrivial and have not previously been remarked-up on or examined. 

\subsection{Derivation from geodesic deviation}

A second type of derivation, based on the Eq.\ of geodesic deviation \cite{wald:1984:gr}, can be found in the primary literature description of interferometric detectors \cite{forward:1978:wlg} \cite[Eqs~2.4--2.8]{schutz:1987:apo}, textbooks \cite[Eqs.~9.24--9.28]{schutz:1990:fci}, and review articles \cite[Eqs.~2.12--2.14]{schutz:2001:gws}. In this derivation one identifies a smooth, one-parameter family of geodesics $\gamma_\lambda(t)$: i.e., for each $\lambda\in\mathbb{R}$, $\gamma_\lambda(t)$ is a geodesic with affine parameter $t$. The vector field $\zeta^\mu(\lambda)=\partial/\partial\lambda$ connects neighboring geodesics (e.g., the beam splitter world line and the end mirror world line) and also satisfies the Eq.\ of geodesic deviation, 
\begin{equation}\label{eq:gd}
\frac{d^2\zeta^\alpha}{dt^2} = -{R^\alpha}_{\beta\gamma\delta}U^\beta \zeta^\gamma U^\delta = -\frac{1}{2}\eta^{\alpha\mu}\zeta^\gamma\frac{\partial^2\hfill}{\partial t^2}h_{\mu\gamma} + \mathcal{O}(h^2).
\end{equation}
The geodesics $\gamma_\lambda$ are identified to be the coordinate fixed world lines of the interferometer end mirror and beam splitter, and the variations in the length of $\zeta$ are equated to variations in the light travel time between the beam splitter and the end mirror, leading to 
\begin{equation}
\left|\zeta^\mu\zeta_\mu\right|^{1/2} = L\left(1+\frac{1}{2}h_{ij}\zeta^i\zeta^j\right), 
\end{equation}
valid in the limit that $fL\ll1$, where $f$ is the gravitational wave frequency. 

In this argument one makes the following assumptions:
\begin{itemize}
\item The spatial projection of the null-ray separating the nearby geodesics is also the geodesic deviation vector $\zeta^\mu$; 
\item The coordinate speed of light propagation is unity so that the spatial coordinate distance traveled along $\zeta^\mu$ (i.e., $L$) can be equated to the coordinate propagation time. 
\end{itemize}
Both of these assumptions are false at order $\mathcal{O}(h)$. Incorrectly assuming that $\zeta^\mu$ is the spatial projection of the null-ray tangent leads to $\mathcal{O}(h)$ errors in the spatial distance traveled by the null-ray. Incorrectly assuming that the coordinate velocity of light is unity also makes errors of order $\mathcal{O}(h)$.

In a pedagogical article on the importance of relating physical observables only to gauge invariant quantities, Garfinkle \cite{garfinkle:2006:gia} finds, in the limit of slowly varying fields, a second-order differential Eq.\ for the phase difference $\Delta\phi$ in terms of the Riemann tensor. This Eq.\ has the form of the geodesic deviation Eq.\ in this limit; however, derived in the way shown it avoids the questions that arise when the geodesic deviation Eq.\ is invoked from the outset and without argument or justification, and can be correctly generalized to more general circumstances. 

\section{The response of a free-mass, one-pass interferometric gravitational wave detector}\label{sec:correct}
\subsection{Introduction}
In this section we derive the response of a one-pass interferometric gravitational wave detector. While the derivation's details will be specialized for the case of free masses and Minkowski background spacetime we take special care to indicate where the different  specializations are made and how the calculation would proceed in the more general case. The presentation that follows thus serves as a template for more general applications that require similar calculations. 

As described in Sec.~\ref{sec:observable} the observable is the difference in phase $\Delta\phi$ of the electromagnetic wavefronts inbound at the beam splitter. As shown schematically in Figure~\ref{fig:ifo}, the phase of each inbound wavefront is equal to the phase of the electromagnetic wave outbound from the beam splitter at an earlier time. The principal goal in calculating the response of an interferometer is thus calculating the elapsed proper time at the beam splitter between when an electromagnetic wavefront is inbound at the beam splitter and when it was earlier outbound from the beam splitter. If the beam splitter's spacetime trajectory is $\mathsf{z}^\mu(\tau)$, where $\tau$ is the proper time measured along the beam splitter worldline, the elapsed coordinate time $\Delta t$ is readily converted to the elapsed proper time $\Delta\tau$ either by solving the implicit equations 
\begin{subequations}
\begin{equation}
\begin{array}{rcl}
\mathsf{z}^t(\tau) &=& t\\
\mathsf{z}^t(\tau-\Delta\tau) &=& t-\Delta t,
\end{array}
\end{equation}
or by evaluating the integral
\begin{equation}
\Delta\tau = \int_{t-\Delta t}^t dt\,\left(\frac{d\mathsf{z}^t}{d\tau}\right)^{-1}.
\end{equation}
\end{subequations}

Henceforth we concentrate on calculating the coordinate time difference $\Delta t$ separating the emission of a wavefront and its subsequent return to the beam-spliter. We do so in two steps. In the first step we calculate the coordinate time required for an electromagnetic wavefront to propagate from the end mirror to the beam splitter, arriving at time $t$; in the second step, we calculate the coordinate time required for the wavefront to propagate from the beam splitter to the end mirror, arriving at time $t'$. From these the coordinate time of the round-trip can be calculated for each arm and, as described above, the observable phase difference $\Delta\phi(\tau)$. 

\subsection{Inbound phase-front propagation}
Focus attention on an interferometer arm consisting of a beam splitter and an end mirror. An  electromagnetic wavefront arrives at the beam splitter from the end mirror at time coordinate time $t$. Define $\Delta_{\text{i}}(t)$ as the elapsed coordinate time between the event of the wavefront leaving the end mirror and arriving at the beam splitter at time $t$; i.e., if the wavefront arrives at the beam splitter at time $t$ it left the end mirror at coordinate time $t-\Delta_{\text{i}}(t)$. 

\subsubsection{Trajectory}\label{sec:traj}
The trajectory $\mathsf{x}^\mu(\lambda)$ of an electromagnetic wavefront can be evaluated from the geodesic equation:
\begin{subequations}\label{eq:geodesic}
\begin{equation}
0 = \frac{d^2\mathsf{x}^\mu}{d\lambda^2} + \Gamma^{\mu}_{\alpha\beta}\frac{d\mathsf{x}^\alpha}{d\lambda}\frac{d\mathsf{x}^\beta}{d\lambda}
\end{equation}
where $\lambda$ is an affine parameter along the geodesic and the condition that the geodesic be null is 
\begin{equation}\label{eq:null}
0 = g_{\mu\nu}\frac{d\mathsf{x}^\mu}{d\lambda}\frac{d\mathsf{x}^\nu}{d\lambda}. 
\end{equation}
\end{subequations}
We introduce the gravitational wave perturbation by writing the metric as the sum of a background $g^{(0)}_{\mu\nu}$ and a gravitational wave perturbation $\epsilon h_{\mu\nu}$
\begin{equation}
g_{\mu\nu} = g^{(0)}_{\mu\nu} + \epsilon h_{\mu\nu} + \mathcal{O}(\epsilon^2),
\end{equation}
where we are explicit about the perturbation expansion parameter $\epsilon$. 

The first step in our calculation is to solve equations (\ref{eq:geodesic}), subject to the constraint in Eq.\ (\ref{eq:null}), perturbatively about $\epsilon=0$; i.e., writing the trajectory $\mathsf{x}^\mu(\lambda;\epsilon)$ as 
\begin{subequations}\label{eq:expansion}
\begin{align}
\mathsf{x}^\mu &= \sum_{p}\epsilon^p \mathsf{x}_p^\mu(\lambda)
\end{align}
\end{subequations}
we solve equations (\ref{eq:geodesic}) order-by-order in $\epsilon$.

For the particular case of interest the background spacetime is Minkowski (i.e., $g^{(0)}_{\mu\nu} = \eta_{\mu\nu}$), the beam splitter and the end mirror are free and, in the limit $\epsilon\rightarrow0$, at relative rest. We take advantage of this circumstance to express the gravitational wave perturbation in the corresponding TT gauge. In this spacetime and with these coordinates the connection coefficients are given by 
\begin{equation}
\Gamma^{\alpha}_{\beta\gamma} = \frac{\epsilon}{2}\eta^{\alpha\mu}\left[h_{\mu\beta,\gamma}+h_{\mu\gamma,\beta}-h_{\gamma\beta,\mu}\right] + \mathcal{O}(\epsilon^2),
\end{equation}
and the geodesic equations become 
\begin{subequations}
\begin{align}
0 &=  \frac{d^2\mathsf{t}}{d\lambda^2} + \frac{\epsilon}{2}\frac{d\mathsf{x}^l}{d\lambda}\frac{d\mathsf{x}^m}{d\lambda}\frac{\partial h_{lm}}{\partial t}
+\mathcal{O}(\epsilon^2)\\
0 &= \frac{d^2\mathsf{x}^j}{d\lambda^2} 
+\epsilon\frac{d\mathsf{x}^p}{d\lambda}\left[
\delta^{jl}\frac{d\mathsf{t}}{d\lambda}\frac{\partial h_{lp}}{\partial t}
+\delta^{jl}\frac{d\mathsf{x}^m}{d\lambda}\frac{\partial h_{lp}}{\partial x^m}
-\delta^{jm}\frac{1}{2}\frac{d\mathsf{x}^l}{d\lambda}\frac{\partial h_{lp}}{\partial x^m}\right] 
+\mathcal{O}(\epsilon^2),\label{eq:traj}
\end{align}
\end{subequations}
where we have written $\mathsf{t}$ for $\mathsf{x}^t$. 

At order $\epsilon^0$ we have 
\begin{subequations}\label{eq:order0}
\begin{align}
0& = \frac{d^2\mathsf{t}_0}{d\lambda^2}&\frac{d\mathsf{t}_0}{d\lambda}&=\sigma_0&\mathsf{t}_0 &= \sigma_0\left(\lambda-\lambda_0\right)\\
0&= \frac{d^2\mathsf{x}_0^j}{d\lambda^2}&\frac{d\mathsf{x}_0^j}{d\lambda}&= \sigma_0n^j_0& \mathsf{x}_0^j &= (\lambda-\lambda_0)\sigma_0n^j_0+{y}^j_0
\end{align}
where $\lambda_0$, $\sigma_0$, $n^j_0$ and ${y}^j_0$ are all constants of the integration, with $\lambda_0$ and $\sigma_0$ corresponding to the arbitrary zero and scale of the affine parameter $\lambda$, and the $n^j_0$ and $y^j_0$ corresponding to the wavefront propagation direction and location at $\lambda=\lambda_0$. The condition that the trajectory is a null-geodesic is enforced by Eq.\ (\ref{eq:null}), which becomes at $\mathcal{O}(\epsilon^0)$
\begin{equation}
0 =  -1+ n_0^ln_0^m\delta_{lm}
\end{equation}
\end{subequations}

Now turn to the order $\epsilon^1$ trajectory. The order $\epsilon^1$ corrections to the geodesic trajectory satisfy\begin{align}
\frac{d^2\mathsf{t}_1}{d\lambda^2} &= -\frac{\sigma_0^2}{2}n_0^l{}n_0^m\left.\frac{\partial h_{lm}}{\partial t}\right|_0\\
\frac{d^2\mathsf{x}_1^j}{d\lambda^2} &= -\frac{d\mathsf{x}_0^p}{d\lambda}\left.\left[
\delta^{jl}\frac{d\mathsf{t}_0}{d\lambda}\frac{\partial h_{lp}}{\partial t}
+\delta^{jl}\frac{d\mathsf{x}_0^m}{d\lambda}\frac{\partial h_{lp}}{\partial x^m}
-\frac{1}{2}\delta^{jm}\frac{d\mathsf{x}_0^l}{d\lambda}\frac{\partial h_{lp}}{\partial x^m}\right]\right|_0\,,
\end{align}
where $\big|_0$ indicates that 
the partial derivatives of $h_{lm}$ are evaluated along the 0th-order 
trajectory $(\mathsf{t}_0,\mathsf{x}^j_0)$.
For plane wave $h_{lm}$ we have
\begin{subequations}
\begin{align}
h_{lm}(t,{x^j})  &= \mathsf{h}_{lm}(u(t,x))\\
u(t,x) &= t-{k}_jx^j 
\end{align}
where ${k}^l$ is a unit vector in the unperturbed geometry:
\begin{equation}
0 = -1 + k_lk_m\delta^{lm}.
\end{equation}
\end{subequations}
Consequently, 
\begin{subequations}
\begin{align}
\frac{\partial h_{lm}}{\partial t} &=\frac{d \mathsf{h}_{lm}}{du}\frac{\partial u}{\partial t} = \frac{d \mathsf{h}_{lm}}{du}\\
\frac{\partial h_{lm}}{\partial x^j} &=\frac{d \mathsf{h}_{lm}}{du}\frac{\partial u}{\partial x^j} = -k_j\frac{d \mathsf{h}_{lm}}{du}.
\end{align}
Writing for the metric perturbation along the wavefront
\begin{equation}
{H}_{lm}(\lambda) = \mathsf{h}_{lm}\left(u(\mathsf{t}(\lambda),\mathsf{x}(\lambda))\right)= h_{lm}\left(\mathsf{t}(\lambda),\mathsf{x}^j(\lambda)\right) .
\end{equation}
we have
\begin{equation}
\frac{d \mathsf{h}_{lm}}{du}
=\frac{d{H}_{lm}}{d\lambda}\left[\frac{d}{d\lambda}u(\mathsf{t}(\lambda),\mathsf{x}^j(\lambda))\right]^{-1}
\end{equation}
Invoking the $\mathcal{O}(\epsilon^0)$ trajectory (eqs. \ref{eq:order0}) we find:
\begin{equation}
\frac{d}{d\lambda}u(\mathsf{t}(\lambda),\mathsf{x}^j(\lambda)) = \frac{\partial u}{\partial t}\frac{d\mathsf{t}}{d\lambda} + \frac{\partial u}{\partial x^j}\frac{d\mathsf{x}^j}{d\lambda}= \sigma_0\left(1-k_jn_0^j\right)+\mathcal{O}(\epsilon). 
\end{equation}
Thus we may write
\begin{align}
\frac{\partial h_{lm}}{\partial t} &= \frac{1}{\sigma_0\left(1-k_jn_0^j\right)}\frac{d{H}_{lm}}{d\lambda} + \mathcal{O}(\epsilon)\\
\frac{\partial{h_{lm}}}{\partial x^j} &=
-
\frac{k_j}{\sigma_0\left(1-k_pn^p _0\right)}
\frac{d{H}_{lm}}{d\lambda} + \mathcal{O}(\epsilon).
\end{align}
\end{subequations}

For plane wave $h_{lm}$ we thus have
\begin{subequations}
\begin{align}
\frac{d\mathsf{t}_1}{d\lambda} &= \sigma_0\left.\left[\sigma_1 -\frac{1}{2}\frac{{}n_0^l{}n_0^m}{1-{k_jn_0^j}}{H}_{lm}\right]\right|_0\\
\mathsf{t}_1(\lambda) &= \sigma_0\left[\sigma_1(\lambda-\lambda_0) -\frac{1}{2}\frac{{}n_0^l{}n_0^m}{1-{k_jn_0^j}}\mathcal{H}_{lm}(\lambda_0,\lambda,\mathsf{t}_0,\mathsf{x}_0^j)\right]\\
\frac{d\mathsf{x}^j_1}{d\lambda} &= \sigma_0\left.\left[n^j_1
-\left(\delta^{jl}
+\frac{1}{2}\frac{n_0^l\delta^{jm}k_m}{1-k_pn_0^p}
\right)n_0^p{H}_{lp}\right]\right|_0\\
\mathsf{x}^j_1(\lambda) &= y^j_1 + \sigma_0\left[(\lambda-\lambda_0){n}_1^j- \left(
\delta^{jl}+\frac{1}{2}\frac{n_0^l\delta^{jm}{k}_m}{1-{k_pn_0^p}}
\right)n_0^p\mathcal{H}_{lp}(\lambda_0,\lambda,\mathsf{t}_0,\mathsf{x}_0^j)\right]
\end{align}
where $y^j_1$, $n_1^j$ and $\sigma_1$ are constants of the integration and we have defined the functional $\mathcal{H}$ of the trajectory $\{\mathsf{t}(\lambda),\mathsf{x}^j(\lambda)\}$ as
\begin{equation}
\mathcal{H}_{lm}(\lambda_0,\lambda,\mathsf{t},\mathsf{x}^j)=\int^\lambda_{\lambda_0} d\lambda'\,{H}_{lm}(\lambda') = \int_{\lambda_0}^\lambda d\lambda'\,h_{lm}(\mathsf{t}(\lambda'),\mathsf{x}^j(\lambda')).
\end{equation}
\end{subequations}

Finally, the condition that the trajectory be a null-geodesic becomes
\begin{subequations}
\begin{align}
0&=
-2\frac{d\mathsf{t}_0}{d\lambda}\frac{d\mathsf{t}_1}{d\lambda}
+2\frac{d\mathsf{x}_0^l}{d\lambda}\frac{d\mathsf{x}_1^m}{d\lambda}\delta_{lm}
+\frac{d\mathsf{x}_0^l}{d\lambda}\frac{d\mathsf{x}_0^m}{d\lambda}h_{lm}(\mathsf{t}_0,\mathsf{x}^j_0)
\nonumber\\%
&= \left(n_0^ln_1^m\delta_{lm}-\sigma_1\right)\sigma_0^2. 
\end{align}
\end{subequations}

\subsubsection{Renormalization}\label{sec:renorm}

It will in general be the case that $\epsilon\mathsf{t}_1$ and $\epsilon\mathsf{x}_1^j$ involve terms that are unbounded in $\lambda$. For example, in the calculation presented here the terms in question are those linear in $(\lambda-\lambda_0)$. Owing to this unbounded growth the perturbation expansion breaks down when $(\lambda-\lambda_0)\epsilon$ is no longer small. These terms are generally evidence that ``constants of the integration'' introduced in solving for $\mathsf{t}_0$ and $\mathsf{x}^j_0$ should no longer be considered constants at $\mathcal{O}(\epsilon)$. The next step in determining the null-geodesic trajectory of an electromagnetic wavefront is to regularize any divergent terms in the perturbation expansion for $\mathsf{t}$ and $\mathsf{x}^j$. This is generally done using renormalization group methods \cite{chen:1994:rgt,chen:1996:rga}.  
In our particular case (Minkowski space and TT gauge perturbation) the renormalization is particularly simple: introducing the renormalized $\sigma$,  $\mathsf{n}^j$, and $\mathsf{y}^j$,  
\begin{subequations}
\begin{align}
\sigma_0 = \sigma(1-\epsilon \sigma_1)\label{eq:a}\\
\sigma_0n_0^j = \sigma(\mathsf{n}^j - \epsilon n_1^j)\\
y_0^j = \mathsf{y}^j - \epsilon y_1^j,
\end{align}
\end{subequations}
the renormalized perturbed null-geodesic trajectory becomes
\begin{subequations}\label{eq:nullTraj}
\begin{align}
\frac{d\mathsf{t}}{d\lambda} &= \sigma\left[1 - \frac{\epsilon}{2}\frac{\mathsf{n}^l\mathsf{n}^m}{1-k_j\mathsf{n}^j}{H}_{lm}(\lambda)\right] + \mathcal{O}(\epsilon^2)\\
\mathsf{t} &= \sigma\left[\left(\lambda-\lambda_0\right)
- \frac{\epsilon}{2}\frac{\mathsf{n}^l\mathsf{n}^m}{1-k_q\mathsf{n}^q}
\mathcal{H}_{lm}(\lambda_0,\lambda,\mathsf{t}_0,\mathsf{x}_0^j)\right]
+ \mathcal{O}(\epsilon^2)\label{eq:t}\\
\frac{d\mathsf{x}^j}{d\lambda} &=\sigma\left[\mathsf{n}^j 
- \epsilon \left(\mathsf{n}^l\delta^{jm}+\frac{1}{2}\frac{\mathsf{n}^l\mathsf{n}^m}{1-k_j\mathsf{n}^j}\delta^{jp}k_p\right){H}_{lm}(\lambda)
\right]
+\mathcal{O}(\epsilon^2)\label{eq:dx/dl}\\
\mathsf{x}^j &= \mathsf{y}^j+\sigma\left[\left(\lambda-\lambda_0\right)\mathsf{n}^j
-\epsilon{}\left(\mathsf{n}^k\delta^{jl}+\frac{1}{2}\frac{\mathsf{n}^k\mathsf{n}^l}{1-k_q\mathsf{n}^q}\delta^{jm}k_m\right)\mathcal{H}_{kl}(\lambda_0,\lambda,\mathsf{t}_0,\mathsf{x}_0^j)\right]+\mathcal{O}(\epsilon^2)
\label{eq:x(l)}
\end{align}
and the condition that the geodesic be null becomes
\begin{align}\label{eq:nullCond}
%0 &= -\left(\frac{dt}{d\lambda}\right)^2
%+\left(\delta_{lm}+\epsilon h_{lm}\right)\frac{dx^l}{d\lambda}\frac{dx^m}{d\lambda}\\
0&= -1+\mathsf{n}^l\mathsf{n}^m\delta_{lm}
+ \mathcal{O}(\epsilon^2).
\end{align}
\end{subequations}
With this renormalization the perturbation of the null-geodesic trajectory is seen to always remain small. 

\subsubsection{Coordinate velocity along the null trajectory}
Equations (\ref{eq:nullTraj}) give the geodesic null-ray trajectory followed by the electromagnetic wavefront as it propagates between the end mirror and the beam splitter. As suggested in Sec.~\ref{sec:classic} it is \emph{not} the unperturbed trajectory $(\lambda,\lambda\zeta^j)$; rather, it involves $\mathcal{O}(\epsilon^1)$ corrections in both magnitude and direction when compared to that coordinate trajectory. Additionally, the coordinate velocity,
\begin{equation}
\frac{d\mathsf{x}^j}{d\mathsf{t}} = 1-\epsilon\mathsf{n}^l\mathsf{n}^mH_{lm} + \mathcal{O}(\epsilon^2),
\end{equation}
also involves $\mathcal{O}(\epsilon)$ corrections. There is no \emph{a priori} reason to expect that the contribution 
\begin{equation}
\int dt\sqrt{\delta_{ij}\frac{d\mathsf{x}^{i}}{dt}\frac{d\mathsf{x}^{j}}{dt}}
\end{equation} 
to the integral presented in Eq.\ (\ref{eq:rhs}) should evaluate to $L+\mathcal{O}(\epsilon^2)$, as is glibly assumed in the classic derivations when the $\mathcal{O}(\epsilon)$ corrections to the spatial projection of the null-ray, and the $\mathcal{O}(\epsilon)$ corrections to the coordinate velocity, are ignored without comment.

\subsubsection{end mirror and beam splitter worldlines}
The null-ray geodesic trajectory described above intersects the worldlines of the end mirror and beam splitter at coordinate times $t_i$ and $t_f$. Our immediate goal is to determine $\Delta_{\text{i}}(t_f)=t_f-t_i$. To achieve this goal we must know the beam splitter and end mirror worldlines to $\mathcal{O}(\epsilon^2)$. We can solve for these worldlines using the same methodology as we used in Sec.~\ref{sec:traj}--\ref{sec:renorm} to find the null-geodesic trajectory of the electromagnetic wavefronts. Now, however, the equations to be solved for the end mirror or beam splitter worldline $\mathsf{z}$ are
\begin{equation}
\mathsf{a}^\mu = \frac{d^2\mathsf{z}^\mu}{d\tau^2} + \Gamma^{\mu}_{\alpha\beta}\frac{d\mathsf{z}^\alpha}{d\tau}\frac{d\mathsf{z}^\beta}{d\tau}
\end{equation}
where $\mathsf{a}^\mu$ is the four-acceleration acting on the end mirror or beam splitter and boundary conditions are specified on $\mathsf{z}^\mu$ and $d\mathsf{z}^\mu/d\tau$ at some initial moment of time.

In the present case --- Minkowski background, TT gauge perturbation, beam splitter and end mirror free and initially at coordinate rest  --- the four-acceleration $\mathsf{a}^\mu$ vanishes and $d\mathsf{z}^\mu/d\lambda= \delta^\mu_t$ at the initial moment of time. The resulting (geodesic) equations for the beam splitter and end mirror worldlines have the trivial solution
\begin{align}
\mathsf{z}^j(\tau) &= z^j + \mathcal{O}(\epsilon^2)
\end{align}
where $z^j$ is the beam splitter or end mirror location at the initial moment of time. 

\subsubsection{Coordinate light travel-time}

Given the $\mathcal{O}(\epsilon^2)$ worldlines of the electromagnetic wavefront $(\mathsf{t},\mathsf{x}^j)$, the beam splitter $\mathsf{z}_{\text{BS}}^j(t)$ and end mirror $\mathsf{z}^j_{\text{EM}}(t)$ in the perturbed space time we can calculate the coordinate time interval $\Delta_i(t)$ between the events corresponding to the intersection of the null-geodesic $(\mathsf{t},\mathsf{x}^j)$ with the two worldlines $\mathsf{z}^j_{\text{BS}}$ and $\mathsf{z}^j_{\text{EM}}$ by solving the implicit equations
\begin{subequations}\label{eq:DeltaI}
\begin{align}
\mathsf{t}(\lambda_f) &= t\\
\mathsf{z}^j_{\text{BS}}(t) &= \mathsf{x}^j(\lambda_f)\\
\mathsf{z}^j_{\text{EM}}(\mathsf{t}(\lambda_i)) &= \mathsf{x}^j(\lambda_i).
\end{align}
\end{subequations}
The interval $\Delta_i(t)$ is equal to $t-\mathsf{t}(\lambda_i)$. The eight equations (\ref{eq:DeltaI}) and (\ref{eq:nullCond}) determine the two constants $\lambda_f$, $\lambda_i$ and the six constant $\mathsf{y}^j$, $\mathsf{n}^j$ in Eq.\ (\ref{eq:x(l)}) for $\mathsf{x}^j(\lambda)$.

In our particular case $\mathsf{z}^j_{\text{BS}}$ and $\mathsf{z}^j_{\text{EM}}$ are constant to $\mathcal{O}(\epsilon^2)$. Without loss of generality we can orient the coordinates so that the beam splitter is at the origin and the end mirror is at $z^j$.  We introduce constants $L$ and $\zeta^j$, 
\begin{subequations}
\begin{align}
\zeta^j &= \frac{z^j}{L}\\
0 &= -1 + \zeta^l\zeta^m\delta_{lm},
\end{align}
\end{subequations}
which are independent of $\epsilon$. In the absence of the perturbation $L$ is the physical separation, at constant $t$, between the beam splitter and the end mirror.  Finally, we choose the constant affine parameter scale factor $\sigma$ [cf.~Equation \ref{eq:a}] to be unity. 

Consider a null-geodesic emerging from the end mirror at $\lambda_0=0$ and traveling in the direction 
\begin{equation}
\frac{dx^j}{d\lambda}(0)=  \mathsf{n}^j 
- \epsilon \left(\mathsf{n}^l\delta^{jm}+\frac{1}{2}\frac{\mathsf{n}^l\mathsf{n}^m}{1-k_j\mathsf{n}^j}\delta^{jp}k_p\right){H}_{lm}(0)
+\mathcal{O}(\epsilon^2).
\end{equation}
If the wavefront is to arrive at the beam splitter at affine parameter $\lambda$ then $\mathsf{n}^j$ must satisfy
\begin{subequations}
\begin{equation}\label{eq:z(l)}
0 ={z}^j + \lambda \mathsf{n}^j
-\epsilon\left(\mathsf{n}^l\delta^{jm}+\frac{1}{2}\frac{\mathsf{n}^l{}\mathsf{n}^m}{1-{k_q\mathsf{n}^q}}k_p\delta^{jp}
\right)\mathcal{H}_{lm}(0,\lambda,t_0,x^j_0) + \mathcal{O}(\epsilon^2),
\end{equation}
When $\epsilon=0$, $\mathsf{n}^j$ is equal to $-\zeta^j$ and $\lambda$ is equal to $L$. As $\epsilon$ is increased $\mathsf{n}^j(\epsilon)$ and $\lambda(\epsilon)$ change. Noting, however, that $\mathsf{n}^j$ is always  $-\zeta^j+\mathcal{O}(\epsilon)$ we have
\begin{equation}
0 ={z}^j + \lambda \mathsf{n}^j
+\epsilon\left(\zeta^l\delta^{jm}-\frac{1}{2}\frac{\zeta^l{}\zeta^m}{1+{k_q\zeta^q}}k_p\delta^{jp}
\right)\mathcal{H}_{lm}(0,L,t_0,x^j_0) + \mathcal{O}(\epsilon^2).
\end{equation}
\end{subequations}
Contracting this expression with $\delta_{jl}\mathsf{n}^l$ and $\delta_{jl}\zeta^l$ we find 
\begin{subequations}
\begin{align}
0 &=  L\zeta^l\mathsf{n}^m\delta_{lm} + \lambda
- \epsilon\left(1-\frac{1}{2}\frac{{k_j\zeta^j}}{1+{k_q\zeta^q}}\right)
{}\zeta^l{}\zeta^m \mathcal{H}_{lm}(0,L,t_0,x^j_0) + \mathcal{O}(\epsilon^2)\\
0 &= L + \lambda\delta_{lm}\zeta^l \mathsf{n}^m 
+\epsilon\left(1-\frac{1}{2}\frac{k_p\zeta^p}{1+{k_q\zeta^q}}
\right)\zeta^l\zeta^m\mathcal{H}_{lm}(0,L,t_0,x^j_0) + \mathcal{O}(\epsilon^2)
\end{align}
\end{subequations}
which together yield
\begin{equation}
\lambda = L + \epsilon\left(1-\frac{1}{2}\frac{k_p\zeta^p}{1+{k_q\zeta^q}}
\right)\zeta^l\zeta^m\mathcal{H}_{lm}(0,L,t_0,x^j_0) +\mathcal{O}(\epsilon^2).
\end{equation}

Combined with our earlier expression for $\mathsf{t}(\lambda)$ [cf.~Equation \ref{eq:t}] we find that the light reaching the beam splitter at coordinate time $t$ left the end mirror at coordinate time $\Delta _{\text{i}}(t)$ earlier, where
\begin{subequations}
\begin{align}
\Delta_{\text{i}}(t) &= L +\frac{\epsilon}{2}\zeta^l\zeta^m\mathcal{H}_{lm}(t-L,t,\mathsf{t}_0,\mathsf{x}_0^j) + \mathcal{O}(\epsilon^2)\\
\mathsf{t}_0(\lambda) &= \lambda\\
\mathsf{x}_0^j(\lambda) &= \zeta^j(t-\lambda).
\end{align}
\end{subequations}

\subsubsection{Propagation path and angle-of-arrival scintillation}
In the classic derivation the light ray is assumed to propagate along the coordinate straight line path between the beam splitter and the end mirror. From Eqs.~(\ref{eq:x(l)}) and (\ref{eq:z(l)}) we see that the spatial coordinate followed by the light ray emerging from the end mirror at affine parameter $\lambda=0$ is the ``wavy'' line
\begin{subequations}
\begin{equation}
\mathsf{x}^j(\lambda) = z^j+\lambda\mathsf{n}^j
+\epsilon\left(\zeta^k\delta^{jl}-\frac{1}{2}\frac{\zeta^k\zeta^l}{1+k_q\zeta^q}\delta^{jm}k_m\right)
{\mathcal{H}_{kl}(0,\lambda,\mathsf{t}_0,\mathsf{x}_0^j)}+\mathcal{O}(\epsilon^2)
\end{equation}
where
\begin{align}
\mathsf{n}^j &= -\zeta^j
+\epsilon\left(\delta^{jp}-\zeta^j\zeta^p\right)
\left(\delta_p^m
+\frac{1}{2}\frac{\zeta^mk_p}{1+{k_q\zeta^q}}
\right)\frac{\zeta^l\mathcal{H}_{lm}(0,L,\mathsf{t}_0,\mathsf{x}_0^j)}{L} + \mathcal{O}(\epsilon^2).
\end{align}
\end{subequations}
A corollary is that the arrival direction of the inbound null ray is time dependent; i.e., the passing gravitational wave causes the distant source of light to undergo angle-of-arrival ``twinkling''.

\subsection{Outbound light propagation time}
In the same way that we obtained the coordinate time $\Delta_{\text{i}}(t)$ required for a null-ray to propagate from the end mirror and arrive at the beam splitter at time $t$, we can find the coordinate time $\Delta_{\text{o}}(t')$ required for a null-ray to propagate from the beam splitter and arrive at the end mirror at $t'$. An electromagnetic wavefront arriving at the end mirror at time $t$ left the beam splitter at time $t-\Delta_{\text{o}}(t)$, where
\begin{subequations} 
\begin{align}
\Delta_{\text{o}}(t) &=
L +
\frac{\epsilon}{2}
\zeta^l\zeta^m
\mathcal{H}_{lm}(t-L,t,\mathsf{t}_0,\mathsf{x}_0^j) + \mathcal{O}(\epsilon^2)
\end{align}
and the path along which we calculate $\mathcal{H}_{lm}$ is now
\begin{align}
\mathsf{t}_0(\lambda) &= \lambda\\
\mathsf{x}_0^j(\lambda) &= \zeta^j(\lambda+L-t).
\end{align}
\end{subequations}
for $t-L\leq\lambda\leq t$.

\subsection{Round-trip light travel time}
Light arriving at the beam splitter at time $t$ will have left the beam splitter at an earlier time $t-\Delta_{\text{RT}}(t)$, where
\begin{align}
\Delta_{\text{RT}}(t) &= \Delta_{\text{i}}(t) + \Delta_{\text{o}}(t-\Delta_{\text{i}}(t))\nonumber\\
&= \Delta_{\text{i}}(t) + \Delta_{\text{o}}(t-L)+ \mathcal{O}(\epsilon^2).
\end{align}
Consider the monochromatic gravitational plane wave
\begin{equation}
\mathsf{h}_{lm}(u) = e_{lm}(\hat{k})\Re\left[e^{2\pi i fu}\right]
\end{equation}
where $u=t-k_jx^j$,
${k_j}$ is the gravitational wave propagation direction and 
$e_{lm}(k_j)$ is the wave's polarization vector.
The inbound (to the beam splitter) light propagation time is 
\begin{subequations}
\begin{align}
\Delta_{\text{i}}(t) &= L +\frac{ \epsilon}{2}\zeta^l\zeta^m\mathcal{L}^{(\text{i})}_{lm}(t) + \mathcal{O}(\epsilon^2)\\
\mathcal{L}^{(\text{i})}_{lm}(t) &= e_{lm}(k_j)\Re\int_{t-L}^t d\lambda'\,
\exp\left[
2\pi i f\left(\lambda'+k_j\zeta^j(\lambda'-t)\right)
\right]\nonumber\\
&= e_{lm}(k_j)L\sinc\left(\pi fL(1+k_j\zeta^j)\right)
%-kzt+(1+kz)(t-L/2) = t - L +(1-kz)L/2
\Re\left[e^{2\pi i f\left[t-L+L(1-k_j\zeta^j)/2\right]}\right]\,.
\end{align}
\end{subequations}
The outbound light propagation time is 
\begin{subequations}
\begin{align}
\Delta_{\text{o}}(t-L) &= L -\frac{ \epsilon}{2}\zeta^l\zeta^m\mathcal{L}_{lm}^{(\text{o})}(t-2L,t-L,k_j,\zeta^j) + \mathcal{O}(\epsilon^2)\\
\mathcal{L}_{lm}^{(\text{o})}(t-L) &= e_{lm}(k_j)\int_{t-2L}^{t-L} d\lambda'\,
\exp\left[
2\pi i f\left(\lambda'-k_j\zeta^j(\lambda'+2L-t)\right)
\right] \nonumber\\
&= L\sinc\left(\pi f L(1-k_j\zeta^j)\right)
%kzt-2kzL+(1-kz)(t-3L/2) = t-2kzL-(1-kz)3L/2 = t - L + L - 2kzL - (1-kz)3L/2 = t - L - (1+kz)L/2
\Re\left[e^{2\pi if\left[t-L-L(1+k_j\zeta^j)/2\right]}\right]e_{lm}(k_j).
\end{align}
\end{subequations}
The gravitational wave correction to the round-trip light propagation time along arm $\zeta^j$ for light arriving at the beam splitter at time $t$ is thus
\begin{align}\label{eq:delta}
\Delta_{\zeta}(t) - 2L &=\frac{\epsilon}{2}\left[\mathcal{L}_{lm}^{(\text{i})}(t)+\mathcal{L}_{lm}^{(\text{o})}(t-L)\right]\zeta^l\zeta^me_{lm}(k_j).
\end{align}

\subsection{Interferometer response}
Now consider the two arms of an interferometer, $\zeta^j$ and $\xi^k$. The response of the interferometer is defined as half the difference in round trip light travel times for light arriving at the beam splitter along the two arms:
\begin{subequations}
\begin{align}
R(t,\zeta^j,\xi^k,e_{lm},k_l) &= \frac{\Delta_{\zeta}(t)-\Delta_{\xi}(t)}{2},
\end{align}
where $\Delta_\zeta$ and $\Delta_\xi$ are given by Eq.~(\ref{eq:delta}).
When the round trip light travel time in the detector arms is short compared to the period of the gravitational waves --- i.e., $fL\ll1$ --- then we are in the small antenna limit. In this limit the response becomes
\begin{equation}
R_{\text{SA}}(t,\zeta^j,\eta^k,e_{lm},k_l) = 
\epsilon \frac{L}{2}\left(
\zeta^l\zeta^m
-\xi^l\xi^m
\right)e_{lm}(k_j)\cos2\pi ft + \mathcal{O}(\epsilon^2,fL).
\end{equation}
\end{subequations}
These expressions are exactly those found in the ``classic'' derivation.

Even though neither of the two most important intermediate steps in the calculation of the response agree with the ``classic'' derivation --- i.e., the spatial projection of the geodesic null-ray between the two mirrors is not the unperturbed trajectory, and the distance traveled along this path is not the unperturbed distance --- the combined effect of these two errors cancels at $\mathcal{O}(\epsilon)$ leading to the same result as the corrected calculation described above.  The reason for this is the subject of the next section. 

\section{Discussion}\label{sec:discuss}

\subsection{Introduction}

It might seem a remarkable coincidence that 
\begin{itemize}
\item The $\mathcal{O}(\epsilon)$ correction to the spatial projection of the geodesic path and
\item The $\mathcal{O}(\epsilon)$ correction to the spatial coordinate velocity along the geodesic path
\end{itemize}
exactly cancel. To understand the conditions under which this cancellation occurs and the classic derivation yields correct results we walk through the classic derivation in its most general form, filling-in the implicit, but unremarked, requirements necessary for its correctness. 

\subsection{Spatial path predicts coordinate-time}\label{sec:dtdx}

The first significant implicit assumption in the classic derivation is that the coordinate-time required to traverse a null trajectory $(\mathsf{t}_0,\mathsf{x}^j_0)$ in the unperturbed spacetime can be expressed as a function of the spatial path $\mathsf{x}^j_0$ alone. Thus, the classic derivation begins by expressing infinitesimal elements of a null path as a relation between coordinate-time and coordinate space displacements. Expressing the general line-element as
\begin{subequations}
\begin{align}
ds^2 &= -\left(\alpha^2 - \beta_j\beta_k\gamma^{jk}\right) dt^2 + 2\beta_j dx^j dt + \gamma_{ij}dx^idx^j,
\end{align}
where the lapse $\alpha$, shift $\beta_j$ and spatial metric $\gamma_{jk}$ are functions of $(t,x^j)$,
the relation between coordinate-time and space displacements for null paths is 
\begin{align}
dt &= 
\frac{\beta_j}{\alpha^2-\beta^2}{dx^j} + 
\frac{1}{\sqrt{\alpha^2-\beta^2}}\sqrt{\left(\gamma_{jk}+\frac{\beta_j\beta_k}{\alpha^2-\beta^2}\right){dx^j}{dx^k}}. 
\end{align}
where $\gamma^{jk}$ and $\beta^2$ are defined by 
\begin{align}
\delta^j_l &= \gamma^{jk}\gamma_{kl} \\
\beta^2 &= \beta_j\beta_k\gamma^{jk}.
\end{align}
\end{subequations}
Writing the metric coefficients $\alpha$, $\beta_j$ and $\gamma_{jk}$ as the sum of a background and a perturbation,
\begin{subequations}
\begin{align}
\alpha &= \alpha_{(0)} + \epsilon \alpha_{(1)}\\
\beta &= \beta_{(0)j}+\epsilon\beta_{(1)j}\\
\gamma_{jk} &= \gamma_{(0)jk} + \epsilon\gamma_{(1)jk},
\end{align}
the relation between coordinate-time and space displacements for null paths may be written
\begin{align}
dt &= 
\left[
\frac{\beta_{(0)j}}{\alpha_0^2-\beta_0^2}\frac{dx^j}{d\lambda} + 
\frac{1}{\sqrt{\alpha_0^2-\beta_0^2}}\sqrt{\left(\gamma_{(0)jk}+\frac{\beta_{(0)j}\beta_{(0)k}}{\alpha_0^2-\beta_0^2}\right)\frac{dx^j}{d\lambda}\frac{dx^k}{d\lambda}}\right.\nonumber\\
&\left.\qquad{}+\epsilon\left(B_j \frac{dx^j}{d\lambda} + G_{jk}\frac{dx^j}{d\lambda}\frac{dx^k}{d\lambda}\sqrt{\left(\gamma_{(0)lm}+\frac{\beta_{(0)l}\beta_{(0)m}}{\alpha_0^2-\beta_0^2}\right)\frac{dx^l}{d\lambda}\frac{dx^m}{d\lambda}}\right)+\mathcal{O}(\epsilon^2)\right]d\lambda
\label{eq:dt}
\end{align}
where the $B_j$ and $G_{jk}$ are linear in the $\alpha_{(1)}$, $\beta_{(1)_j}$ and $\gamma_{(1)jk}$, and
\begin{align}
\beta_{(0)}^2 &= \beta_{(0)j}\beta_{(0)k}\gamma_{(0)}^{jk}\\
\delta_j^k &= \gamma_{(0)jl}\gamma_{(0)}^{lk}.
\end{align} 
\end{subequations}

In the classic derivation one implicitly assumes that Eq.\ (\ref{eq:dt}) together with a spatial trajectory $\mathsf{x}^j(\lambda)$ allows us to calculate the elapsed coordinate-time required for a null (electromagnetic) wavefront to travel from spatial coordinate $\mathsf{x}^j(\lambda_i)$ to $\mathsf{x}^j(\lambda_f)$ along $\mathsf{x}^j$: i.e., 
\begin{subequations}
\begin{align}
\mathcal{T}(\mathsf{x}^j,\lambda_i,\lambda_f) &= \mathcal{T}_0(\mathsf{x}^j,\lambda_i,\lambda_f) + \epsilon \mathcal{T}_1(\mathsf{x}^j,\lambda_i,\lambda_f)\\
\mathcal{T}_0(\mathsf{x}^j) &= \int_{\lambda_i}^{\lambda_f}
\left[\frac{\beta_{(0)j}}{\alpha_{(0)}^2-\beta_{(0)}^2}\frac{d\mathsf{x}^j}{d\lambda} + 
\frac{1}{\sqrt{\alpha_{(0)}^2-\beta_{(0)}^2}}\sqrt{\left(\gamma_{(0)jk}+\frac{\beta_{(0)j}\beta_{(0)k}}{\alpha_{(0)}^2-\beta_{(0)}^2}\right)\frac{d\mathsf{x}^j}{d\lambda}\frac{d\mathsf{x}^k}{d\lambda}}\right]d\lambda
\label{eq:mathcalT0}\\
\mathcal{T}_1(\mathsf{x}^j) &= \int_{\lambda_i}^{\lambda_f} \left(B_j \frac{d\mathsf{x}^j}{d\lambda} + G_{jk}\frac{d\mathsf{x}^j}{d\lambda}\frac{d\mathsf{x}^k}{d\lambda}\sqrt{\left(\gamma_{(0)lm}+\frac{\beta_{(0)l}\beta_{(0)m}}{\alpha_{(0)}^2-\beta_{(0)}^2}\right)\frac{d\mathsf{x}^l}{d\lambda}\frac{d\mathsf{x}^m}{d\lambda}}\right)d\lambda
\end{align}
\end{subequations}
For this to be true the $\mathcal{O}(1)$ terms on the right-hand side of Eq.\ (\ref{eq:dt}) must be independent of time; otherwise it would also be necessary to know $t(\lambda)$ (and, thus, $t(\lambda_i)$ and $t(\lambda_f)$).

\subsection{Spatial path perturbations are irrelevant}

The second significant implicit assumption required for the validity of the classic derivation is that the perturbation to the null ray's spatial path must be irrelevant to the computation of the coordinate-time required to traverse the path. This condition is subtle: it is \emph{not} the same as requiring that the spatial path be geodesic: geodesic paths have extremal path \emph{length}. It is also not a statement that can be framed in a gauge invariant way on a spatial hypersurface: it is about the extremality, with regard to spatial path perturbations, of the \emph{coordinate} time required to traverse the path. 

To see how this requirement arises, write the trajectory traveled by the null ray as a function of the perturbation parameter $\epsilon$: 
\begin{align}
\mathsf{x}^j(\lambda) &= \mathsf{x}^j_0(\lambda) + \epsilon\mathsf{x}^j_1(\lambda) + \mathcal{O}(\epsilon^2).
\end{align}
The elapsed coordinate-time required for the null ray to travel from $\mathsf{x}^j(\lambda_i)$ to $\mathsf{x}^j(\lambda_f)$ along $\mathsf{x}^j$ is 
\begin{align}
\mathcal{T}(\mathsf{x}^j) &= \mathcal{T}_0(\mathsf{x}^j_0+\epsilon\mathsf{x}^j_1) + \epsilon\mathcal{T}_1(\mathsf{x}^j_0+\epsilon\mathsf{x}^j_1)\\
&= \mathcal{T}_0(\mathsf{x}^j_0+\epsilon\mathsf{x}^j_1) + \epsilon\mathcal{T}_1(\mathsf{x}^j_0) + \mathcal{O}(\epsilon^2)
\end{align}
where we have suppressed the designation of the end-points $\lambda_i$ and $\lambda_f$. In the classic derivation one asserts 
\begin{equation}
\mathcal{T}(\mathsf{x}^j) = \mathcal{T}(\mathsf{x}^j_0) + \mathcal{O}(\epsilon)^2,
\end{equation} 
or, equivalently,
\begin{equation}
\mathcal{T}_0(\mathsf{x}^j_0+\epsilon\mathsf{x}^j_1) = \mathcal{T}_0(\mathsf{x}^j_0) + \mathcal{O}(\epsilon^2).
\label{eq:extremal}
\end{equation}
Eq.\ (\ref{eq:extremal}) is the condition satisfied by a path $\mathsf{x}^j_0$ when it is an extremum of $\mathcal{T}_0$, which is, in general not a geodesic or related to a geodesic of either the spacetime or the constant coordinate-time spatial hypersurface. 

Under what conditions can we expect that $\mathcal{T}_0$ have an extremum? From Eq.\ (\ref{eq:mathcalT0}) we see that it is necessary that the $\beta_{(0)j}$ should vanish. Additionally, the $\mathcal{O}(1)$ terms on the right-hand side of Eq.\ (\ref{eq:dt}) must be time independent: i.e., the ratio $\gamma_{(0)jk}/\alpha^2_{(0)}$ must be independent of coordinate-time. Finally, the end-points of the perturbation --- $\epsilon\mathsf{x}^j_1(\lambda_i)$ and $\epsilon\mathsf{x}^j_1(\lambda_f)$ --- should be no more than $\mathcal{O}(\epsilon^2)$: i.e., the world lines of the beam splitter and the end mirror should be coordinate stationary. 

\subsection{Conclusion}
The classic derivation of the response of an interferometric detector makes a number of significant, unarticulated and generally unjustifiable assumptions regarding the background spacetime, the background spacetime coordinate system, and the perturbation coordinate gauge that significantly limit its applicability and render it unsuitable as an archetype for understanding or for calculation in more general circumstances. For the classic derivation to yield correct results 
\begin{itemize}
\item The background spacetime metric must be expressible in a coordinate system with vanishing shift and where the ratio of spatial metric to squared lapse is coordinate-time independent; 
\item In this background coordinate system the beam splitter and end mirror world lines must be spatial coordinate stationary (but not necessarily geodesic); 
\item The gravitational wave perturbation must be expressible in a gauge such that the perturbation to the beam splitter and end mirror world line coordinates are $\mathcal{O}(\epsilon^2)$; and
\item The spatial projection of the background spacetime null-geodesic followed by the electromagnetic wavefront between the beam splitter and the end mirror must be an extremum of the \emph{coordinate-time} required to traverse the null ray [cf.~Equation \ref{eq:mathcalT0}]. 
\end{itemize}
These conditions hold for flat background spacetimes in Minkowski coordinates, with the beam splitter and end mirror at coordinate rest, and gravitational wave perturbation expressed in TT gauge, which is the usual case considered. The two most commented upon conditions identified in the classic derivation --- that the mirrors are free and at relative rest, and that the perturbation is in TT gauge --- are in fact neither  necessary nor sufficient conditions for the validity of the classic derivation (though they are consequences of the necessary conditions when applied to the physical conditions of this particular problem).

The subtlety of these conditions becomes apparent when other circumstances, where the classic derivation might be expected to hold, are considered.  For example, consider isotropic cosmological spacetimes. In these spacetimes the gravitational wave perturbation may be expressed in a TT gauge with respect to an end mirror and beam splitter at rest with respect to the cosmological fluid. When the background is expressed in ``arc-time'' coordinates, 
\begin{equation}
ds^2 = a(t)^2\left[-d\eta^2+ \frac{dr^2}{1-kr^2} + r^2\left(d\theta^2+\sin^2\theta\,d\phi^2\right)\right]. 
\end{equation}
all of the conditions described above hold and the standard derivation will provide the correct result; however, if the same physical problem is expressed in the usual Robertson-Walker (synchronous) coordinates, 
\begin{equation}
ds^2 = -dt^2 + a^2(t)\left[\frac{dr^2}{1-kr}+r^2\left(d\theta^2 + \sin^2\theta\,d\phi^2\right)\right],
\end{equation} 
then $\gamma_{jk}/\alpha^2$ is \emph{not} independent of coordinate-time and the classic derivation fails, even though the spatial hypersurfaces and the spatial coordinate expression of the mirror and electromagnetic wavefront world lines are all unaltered. 

In summary, the adage ``measure twice, cut once'' seems particularly apt. It is fortunate that, in this case, the unarticulated and apparently unrealized assumptions made in the classic derivations of the interferometer response cancel. Nevertheless, it is important that errors like these be recognized and corrected in order that future calculations upon which experiment or observation rely do not make similar mistakes in less fortuitous circumstances. 

\acknowledgments 
It is a pleasure to thank Joseph D. Romano for many in-depth discussions and detailed comments on this manuscript, and for drawing my attention to \cite{whelan:2007:hct} and related work in preparation by Rakhmanov. It is also my pleasure to thank David Garfinkle for drawing my attention to \cite{garfinkle:2006:gia}. I gratefully acknowledge the support of National Science Foundation Grant No.~PHY 06-53462 and No.~PHY 05-55615, and NASA Grant No.~NNG05GF71G, awarded to The Pennsylvania State University.

\end{document}